\documentclass[aps,superscriptaddress,twocolumn,amssymb,showpacs,pra]{revtex4-1}
\usepackage{graphicx}
\usepackage{epsfig}
\usepackage{bm}
\usepackage{amsmath}
\usepackage{color}

\begin{document}
\title{Influence of isolated and clustered defects on electronic and dielectric
properties of w\"ustite}

\author{Urszula D. Wdowik  }
\affiliation{Institute of Technology, Pedagogical University,
             Podchor\c{a}\.zych 2, PL-30084 Krak\'{o}w, Poland}
\author{Przemys\l{}aw Piekarz}
\author{Pawe\l{} T. Jochym}
\author{Krzysztof Parlinski}
\affiliation{Institute of Nuclear Physics, Polish Academy of Sciences,
             Radzikowskiego 152, PL-31342 Krak\'ow, Poland}
\author{Andrzej M. Ole\'{s}}
\affiliation{Marian Smoluchowski Institute of Physics, Jagiellonian University, 
              prof. S. \L{}ojasiewicza 11, PL-30348 Krak\'ow, Poland}
\affiliation{Max-Planck-Institut f\"{u}r Festk\"{o}rperforschung, 
             Heisenbergstrasse 1, D-70569 Stuttgart, Germany}

\begin{abstract}

The influence of intrinsic Fe defects in FeO (either single cation 
vacancies or prototypical 4:1 vacancy clusters) on the electronic and 
dielectric properties is studied within the density functional theory.
The importance of local Coulomb interactions at Fe atoms is highlighted 
and shown to be responsible for the observed insulating Mott gap in FeO 
which is reduced by the presence of defects. 
We investigate nonstoichiometric configurations of Fe$_{1-x}$O with $x$ 
ranging from 3 to 9\% and find the aliovalent Fe cations in both the 
regular and interstitial lattice sites of the considered configurations. 
Furthermore, we show that the trivalent Fe ions are induced by both 
isolated and clustered Fe-vacancies and introduce the empty band states 
inside the insulating gap, which decreases monotonically with 
increasing cation vacancy concentration.
The Fe$_{1-x}$O systems with high defect content become 
metallic for small values of the Coulomb interaction $U$, 
yielding the increase in the dielectric functions and optical 
reflectivity at low energies in agreement with the experimental data.
Due to the crystal defects, the infrared-active transverse 
optic phonons split and distribute over a wide range of frequencies
clarifying the origin of the exceptionally 
large spectral linewidths of the dielectric loss functions observed 
for w\"ustite in recent experiments. 

\end{abstract}

\date{\today}


\maketitle

\section{Introduction}

W\"{u}stite, Fe$_{1-x}$O, has challenged experimental and theoretical 
condensed matter physics for over sixty years. In particular, the 
extent of its nonstoichiometry and its defect structure have been the 
subject of numerous experimental studies 
\cite{Roth60,And77,McC84,McC85,McC94} and are still a matter of 
discussion and a source of controversy \cite{Hazen84}. Generally, these 
studies indicate that the crystal lattice of FeO is modified by the 
presence of Fe$^{3+}$ cations and vacant Fe$^{2+}$ sites. The crystal 
structure and charge distribution in Fe$_{1-x}$O are very complex as 
the defects are likely to coalesce and arrange into clusters, which in 
turn can aggregate into larger defect structures. Although the size, 
shape and distribution of the stable equilibrium defect clusters and 
their aggregates are still controversial, they are expected 
to vary with pressure, temperature and composition \cite{Long91}. 
Moreover, the clusters of defects are believed to order magnetically 
below the N\'{e}el temperature $T_N=198$~K.

Early theoretical investigations on nonstoichiometric FeO were mainly 
aimed at the mechanism of formation and stability of isolated Fe 
vacancies, 
prototypical 4:1-type clusters comprised of interstitial (tetrahedral) 
Fe$^{3+}$ ion surrounded by four cation vacancies as well as complexes 
of such clusters \cite{Catlow75,Press87,Kho89,Sch95}. The results of these 
studies indicate that w\"{u}stite favors formation of 4:1-type 
clusters which are major building blocks for further agglomeration and 
defect cluster growth \cite{Roth60}. Nevertheless, a substantial 
concentration of free vacancies can also be stabilized within the 
w\"ustite lattice.  

Extended defect structures in Fe$_{1-x}$O were rarely modeled using the 
density functional theory (DFT). Such simulations are still extremely 
demanding, as they require proper description of the highly correlated 
nature of 3$d$ electrons \cite{Tran06,Rodl12} and extensive ionic 
relaxations within large supercells to predict a correct ground state of 
Fe$_{1-x}$O. 
In our previous work, the DFT investigations have been performed for 
vacancy-defected FeO \cite{Wdowik13}, uncovering pronounced changes of 
the electronic and vibrational properties induced by isolated cation 
vacancies. 
Recently, it was shown that polaronic distributions
of charge resembling those in magnetite Fe$_3$O$_4$
emerge for the most stable defect structures in w\"{u}stite \cite{Ber14}.

The point defects as well as their clusters are likely to 
affect the dielectric and optical properties of FeO. The results of 
various experiments 
\cite{Bowen75,Prevot77,Kugel77,Hof03,Hir09,Seagle09,Sch12,Kant12,Park13} 
indicate some peculiar behavior of the measured optical quantities 
in w\"ustite and assign it to the native defects. 
However, a detailed explanation of the influence of such defects on the 
electronic and dielectric properties of FeO remains unclear. 
Furthermore, sensitivity of both the electronic structure and dielectric 
functions to the defect states as well as the extent of modifications 
induced by isolated or cluster defects still remains unexplored. 

The present work extends description of w\"ustite by 
investigating its electronic structure and dielectric properties as  
functions of the concentration of cation monovacancies and vacancy 
clusters. 
Our theoretical studies take into account strong local interactions 
of Coulomb type to properly describe the electronic properties
and to better understand the mechanisms of the defect-induced 
changes in Fe$_{1-x}$O.

The paper is organized as follows. The details of the calculation 
method are described in Sec.~\ref{sec:method}. Section~\ref{sec:es}
characterizes electronic properties, whereas Sec.~\ref{sec:results} 
and \ref{sec:exp} present theoretical derivation of the dielectric 
properties and comparison with the experimental data, respectively. 
Finally, Sec.~\ref{sec:sum} summarizes the current results 
and provides conclusions.

\section{Methodology}
\label{sec:method}

Calculations were performed within the spin-polarized DFT method using 
the {\sc vasp} code \cite{VASP}. Electron-ion interactions were 
described in the framework of the projector-augmented wave (PAW) method 
\cite{PAW}. Valence electrons of Fe and O atoms were represented by the 
($3d^7 4s^1$) and ($2s^2 2p^4$) configurations, respectively. 
The gradient-corrected exchange-correlation functional 
in the form proposed by Perdew and Wang (GGA-PW91) \cite{GGA} 
together with a plane-wave expansion up to 520~eV were applied. Effects of electron 
correlations beyond the GGA were taken into account within the framework of 
GGA+$U$ and the approach of Dudarev \textit{et al.} \cite{GGA+U}, where 
a single parameter $U_{\mathrm{eff}}=U-J$ determines an orbital-dependent 
correction to the DFT energy with $J=1$~eV denoting the local 
exchange interaction. 
Most of the present calculations were performed 
using $U_{\mathrm{eff}}=5$~eV, 
which is recognized as a realistic value while describing electron 
correlations  between Fe($3d$) states \cite{Tran06,Rodl12,Wdowik13}. 
Furthermore, it produces the lattice constant $a=4.35$~\AA, which 
remains in a good agreement with various experimental studies \cite{McC84,McC85}.
Additional calculations with reduced values of $U_{\mathrm{eff}}$, 
ranging from  2 to 4~eV, were carried out to investigate the influence 
of strong electron interactions and the screening processes due to 
defects on both the electronic structure and dielectric functions of w\"ustite.

A variety of experiments performed over decades indicates that FeO is 
always deficient in iron and the concentration of vacancies in 
the oxygen sublattice is several orders of magnitude smaller than 
the concentration of iron vacancies \cite{Roth60,And77}. 
Following the experimental evidence, we introduced 
vacancies into the cation sublattice of FeO and left its anion 
sublattice defect-free. The Fe$_{1-x}$O structures with vacancy 
concentrations $x=3$, $5$, $6$, $9$\%  were modeled by supercells 
\cite{Wdowik13,Wdowik08,Wdowik11}. 
Each initial supercell of Fe$_{1-x}$O was derived from the basic 
64-atom FeO supercell which takes into consideration the antiferromagnetic 
(AFII) order \cite{Roth58} and comprises two ferromagnetic Fe-sublattices differing 
in the orientation of the spin magnetic moments (spin-up and spin-down 
Fe-sublattices). Neutral cation vacancies were created by removing the 
appropriate number of Fe atoms from the AFII supercell, whereas 
interstitial Fe atoms are inserted into the empty tetrahedral positions 
of such supercell. The  AFII supercell lacking one Fe atom corresponds 
to Fe$_{1-x}$O with $x=3$\%, while that with two Fe atoms absent 
conforms to Fe$_{1-x}$O with $x=6$\%. In the latter case, each of the 
ferromagnetic sublattices contains one Fe vacancy ($V_{\mathrm{Fe}}$). 
There is, however, a number of possible configurations within the 
simulated supercell  with $x=6$\% differing among each other by 
$V_{\mathrm{Fe}}-V_{\mathrm{Fe}}$ distances. For further considerations 
we have selected the configuration minimizing the system energy and 
corresponding to the $V_{\mathrm{Fe}}-V_{\mathrm{Fe}}$ distance of 
1.22$a$. Details of such calculations as well as more comprehensive 
discussion is given in our previous work \cite{Wdowik13}. 

\begin{figure}[t!]
\includegraphics[width=\columnwidth]{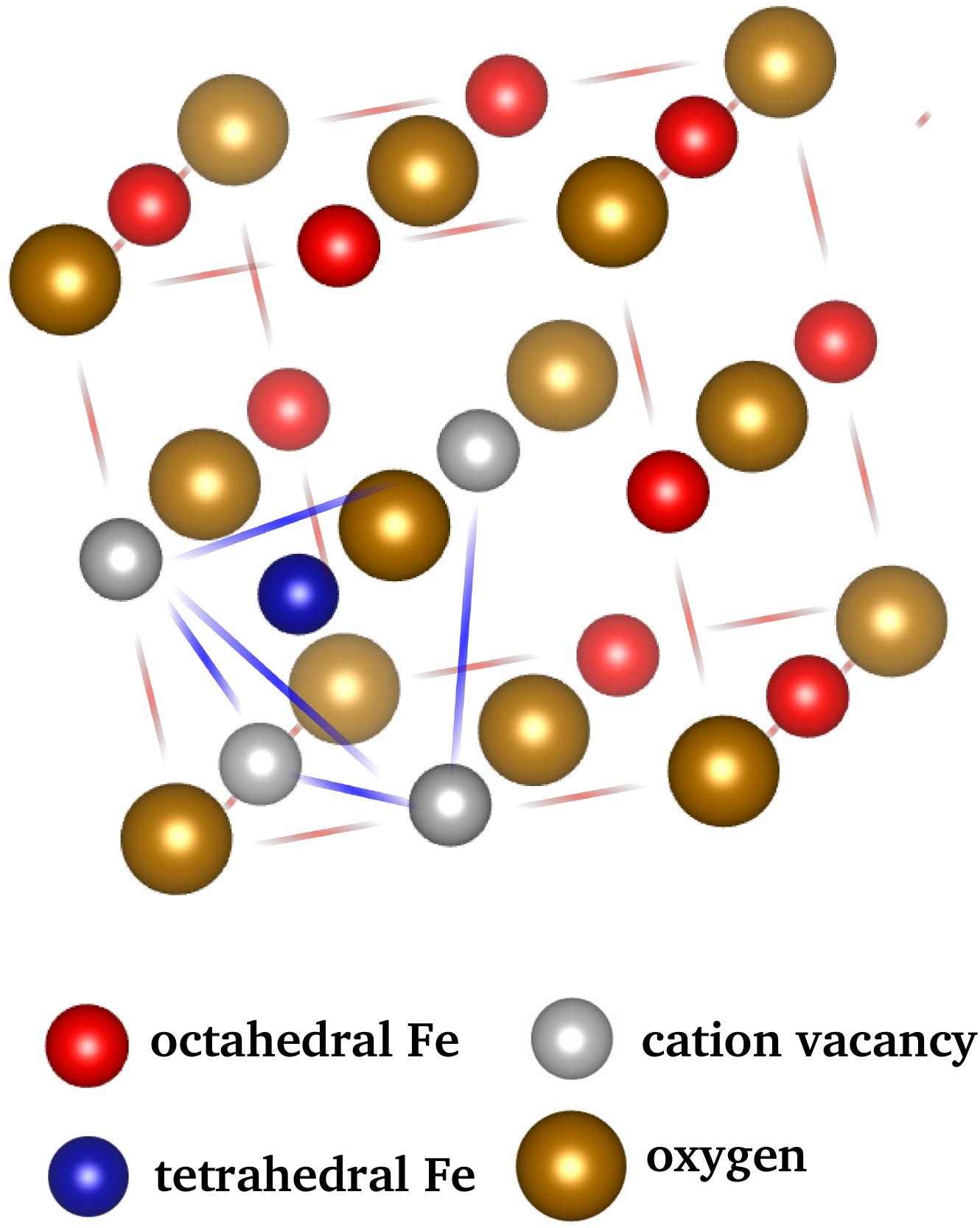}
\caption{(Color online) Schematic representation of the prototypical 
4:1-type cluster with a single Fe ion in a tetrahedral configuration 
of Fe$_{1-x}$O.}
\label{cluster}
\end{figure}  

The Fe$_{1-x}$O structures with $x\simeq 5$\% and $x\simeq 9$\% correspond 
to the 4:1-type clusters. The prototypical 4:1 cluster is schematically 
shown in Fig.~\ref{cluster}. It is composed of an interstitial Fe atom 
surrounded by four cation vacancies arranged into a tetrahedron \cite{Catlow75}. 
We note that due to the AFII magnetic ordering, three cation vacancies 
occupy the spin-up (or spin-down) ferromagnetic sublattice while one cation 
vacancy is always located in the ferromagnetic sublattice with the reversed 
spin direction. The interstitial Fe ion may take on the spin direction of either 
spin-up or spin-down cation sublattice. The AFII supercell containing one 4:1-type 
cluster corresponds to the overall cation deficiency of about 9\%. 
Due to the cubic symmetry of the perfect FeO crystal, all possible locations 
of the 4:1-type cluster in the supercell are equivalent.
The lower concentration of such defects, $x=5$\%, is simulated by placing 
one 4:1 cluster into the supercell elongated in one of the main crystallographic 
axes. For convenience, we summarize the methodology of constructing isolated 
and clustered defects of w\"ustite in Table~\ref{Tab1}.

\begin{table}[t!]
\caption{The list of considered  Fe$_{1-x}$O configurations with the cation
vacancy concentration ($x$), types of defects, sizes of supercells with respect
to the crystallographic unit cell of defect-free FeO, total number of atoms 
in the supercell ($N$), and number of Fe atoms ($N_{\mathrm{Fe}}$),}
\begin{ruledtabular}
\begin{tabular}{ccccc}
 $x$ (\%) & type of defect & supercell size & $N$ & $N_{\mathrm{Fe}}$  \\
\hline
0 & stoichiometric & $2\times2\times2$ & 64 & 32 \\
3 & 1 $V_{\mathrm{Fe}}$  & $2\times2\times2$  & 63 & 31 \\
5 & 4:1-type cluster &  $4\times2\times2$ & 125 & 61 \\
6 & 2 $V_{\mathrm{Fe}}$ & $2\times2\times2$  & 62 & 30 \\
9 & 4:1-type cluster & $2\times2\times2$  & 61 & 29 \\
\end{tabular}
\end{ruledtabular}
\label{Tab1}
\end{table}  

Structural relaxations were performed for fixed volumes of the supercells 
followed from the optimized value of the defect-free FeO lattice. 
This approximation holds for the range of nonstoichiometries 
considered in the present work as the lattice constants of w\"ustite remain
weakly dependent on $x$ \cite{McC84,McC85}. The atomic positions 
were relaxed without symmetry constraints imposed. Only in the case where 
we determine the formation energies of defects, both volumes and atomic positions of 
the supercells were relaxed. However, this procedure did not affect the final results
as the lattice parameters of the supercells containing defects changed by less than 1\%.  
The Brillouin zones were sampled with the $\vec{k}$-point meshes 
of $2\times 2\times 2$ ($x=3, \, 6, \, 9,$\%) and $2\times 2\times 1$ ($x=5$\%) 
generated according to the Monkhorst--Pack scheme. Convergence criteria for the residual
forces and total energies of particular systems were set to 0.01~eV/\AA \ and 0.1~meV, 
respectively. The complex dielectric functions as well as the intensities of 
the infrared-active phonon modes were obtained within the methodology proposed 
by Gajdo\v{s} \textit{et al.} \cite{Gajdos06} and implemented in the {\sc vasp} code.  

The valence charges of Fe ions have been determined from the calculated electron contact 
densities at the Fe nuclei obtained within the full--potential (linearized) augmented 
plane-wave plus local orbitals [FP-(L)APW + lo] method implemented in 
the {\sc wien} code \cite{WIEN}. The wave functions have been expanded 
into spherical harmonics inside 
the nonoverlapping atomic spheres having the radii of muffin-tin sphere $R_{\textrm{MT}}$ 
and in the plane waves within the interstitial region. The $R_{\textrm{MT}}$ 
belonged to the ranges of 1.93--2.03~a.u. for Fe and 1.66--1.74~a.u. 
for O atoms, with a value depending on the simulated system. The maximum 
$l$ value for the expansion of the wave functions into the spherical 
harmonics inside the $R_{\textrm{MT}}$ spheres was set to $l_{max} = 10$, 
while for the expansion of the wave functions within the interstitial 
region the plane-wave cutoff parameter $K_{max}=7/R_{\textrm{MT}}^{min}$
was applied. 
The charge density was Fourier-expanded up to $G_{max} = 12$~Ry$^{1/2}$. 
Here, calculations were performed for the effective Coulomb interaction 
$U_{\mathrm{eff}}=5$~eV, introduced in the rotationally invariant 
form proposed by Anisimov \textit{et al.} \cite{Ani93}. 
Calculations have been carried out for the supercell
volumes adopted from the pseudopotential method calculations. 
Only the atomic position were relaxed within the FP-LAPW methodology
with the force convergence of 0.01 mRy/a.u. and the symmetry constraints removed.

The valence states of Fe ions in Fe$_{1-x}$O systems were identified 
according to the isomer shift ($\delta$) systematics \cite{Shenoy78}. 
In general, the isomer shift is expressed as $\delta=\alpha(\rho-\rho_0)$, 
where $\rho$ and $\rho_0$ stand for the electron contact densities at 
the resonant nucleus in a given matrix and reference material, 
respectively. The symbol $\alpha$ denotes the calibration constant 
characteristic for a particular nuclear transition. Our calculations 
are performed for the 14.41-keV transition in $^{57}$Fe and the isomer 
shifts are given with respect to metallic bcc $\alpha-$Fe upon applying 
previously determined $\alpha=0.291$~(a.u.)$^3$mm/s \cite{Wdowik07}.
More technical details of such calculations can be found in our earlier 
papers \cite{Wdowik07,Wdowik11}.

\section{Electronic structure}
\label{sec:es}

All considered Fe$_{1-x}$O compositions reveal two different types of 
cation valence states, namely high-spin Fe$^{2+}$ and high-spin Fe$^{3+}$ residing both 
in the regular (R) and interstitial (I) positions of the cation sublattice. 
The compositions with 3 and 6\% of isolated vacancies show 
both the Fe$^{2+}$ and Fe$^{3+}$ ions in R sites of the lattice, 
while those having 5 and 9\% of vacancies and corresponding to 
the 4:1-type clusters exhibit Fe$^{2+}$ and Fe$^{3+}$ valence states 
in both the R and I lattice sites. Interstitial Fe becomes divalent 
when its spin is parallel to the spin 
direction of the magnetic sublattice from which three Fe atoms were 
removed, whereas it converts to the trivalent state if its spin is 
antiparallel to this sublattice (i.e., parallel to the ferromagnetic 
sublattice from which one Fe atom was removed). Thus, the interstitial 
Fe$^{2+}$ has the opposite spin 
direction to the interstitial Fe$^{3+}$. 

\begin{figure}[t!]
\includegraphics[width=\columnwidth]{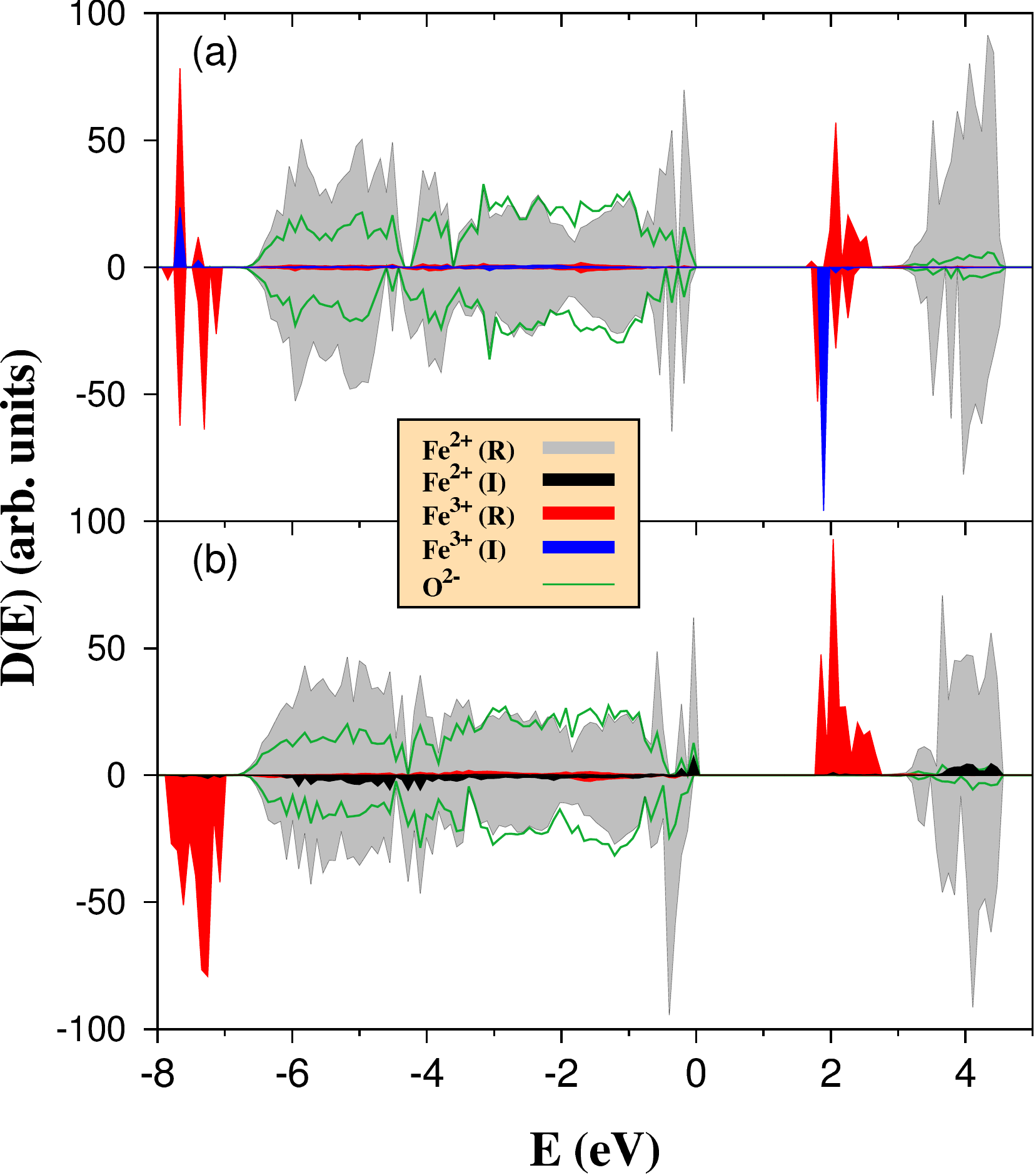} 
\caption{(Color online) Densities of electron states calculated for 
Fe$_{1-x}$O containing 4:1-type vacancy clusters ($x=5$\%) at 
$U_{\mathrm{eff}}=5$~eV. Configuration with: 
(a) interstitial Fe$^{3+}$ ion, and 
(b) interstitial Fe$^{2+}$ ion. 
Positive (negative) values represent the spin-up (spin-down) 
components. The 3$d$ states of Fe$^{2+}$ and Fe$^{3+}$ ions residing 
in the regular (R) and interstitial (I) lattice sites are denoted by 
shaded areas (see the legend). Solid curve corresponds to O$^{2-}$ 
states. 
The top of the valence band is taken as the reference energy (0~eV).}
\label{eDOS}
\end{figure}

The total number of trivalent ions results from the number of cation 
vacancies created in the lattice since each vacancy is accompanied by 
two trivalent configurations. Therefore, six Fe$^{3+}$ ions occupy R 
sites of the w\"ustite cation sublattice containing 4:1-type clusters, 
with Fe$^{2+}$ ion at the I site and five Fe$^{3+}$ ions are located 
in the R lattice positions when the vacancy cluster involves interstitial 
Fe$^{3+}$ ion. The compositions $x=5$\% and $x=9$\% 
with interstitial Fe$^{3+}$ ions have lower energies than the respective
compositions with interstitial Fe$^{2+}$ ions by 14 and 8~meV/atom, 
respectively. This indicates that the w\"ustite lattice favors 
interstitial Fe cations in the trivalent charge state,
while the interstitial Fe$^{2+}$ state can be considered as the metastable state. 
Nevertheless, it is interesting to note essential differences between the electronic 
structures of these two configurations that can be seen in Fig.~\ref{eDOS}. 

First of all, the unoccupied electronic states due to Fe$^{3+}$ ions 
appear in both the spin-up and spin-down channels for the system with 
the interstitial Fe$^{3+}$ ion [Fig.~\ref{eDOS}(a)]. On the other hand, 
these states exist only in one channel direction in the system with 
the interstitially located Fe$^{2+}$ ion [Fig.~\ref{eDOS}(b)]. 
In the latter case, the Fe$^{3+}$ ions induced in the R lattice positions occupy 
solely a single magnetic sublattice, whereas they are created in both 
magnetic sublattices in the former case. The interstitial Fe$^{2+}$ 
ions introduce their electronic states into the same energy bands as 
the Fe$^{2+}$ ions residing in the R lattice sites, as could be 
expected. Therefore, in the Fe$_{1-x}$O systems containing either 4:1 
vacancy clusters or free vacancies \cite{Wdowik13} the bottom of the 
conduction band as well as the states lying at about $-8$ to $-7$ eV below 
the top of the valence band consist of states formed by trivalent Fe 
cations. The empty states in the conduction band located above 
$\sim 3$~eV from the top of the valence band arise from the divalent 
Fe cations occupying the R and I lattice sites. Each Fe$_{1-x}$O 
configuration shows significantly reduced insulating gap by the unoccupied 
electronic bands arising from the Fe$^{3+}$ ions stabilized by cation 
vacancies. 

Relative stability between monovacancies and vacancy clusters in 
Fe$_{1-x}$O has been determined from the respective formation energies 
of neutral vacancies, defined as follows \cite{Eform}, 
\begin{equation}
E_{\mathrm{form}} = E_d - E_{\mathrm{FeO}} + n \, E_{\mathrm{Fe}}, 
\end{equation}
where
$E_d$ is the total energy of Fe$_{1-x}$O supercell containing $n$ vacancies, 
$E_{\mathrm{FeO}}$ is the total energy of the FeO supercell, and 
$E_{\mathrm{Fe}}$ denotes the energy of Fe atom in the metallic Fe bcc 
crystal. It occurs that formation energies of systems with $x=3$ and 
$x=6$\% are respectively higher by 107 and 36~meV per defect
than the formation energy of system containing 4:1-type cluster ($x=9$\%). 
Thus, the w\"ustite lattice seems to prefer formation of vacancy clusters over 
the isolated cation vacancies. This result corresponds very well to the recent 
findings presented in Ref. \cite{Ber14}. 

In principle, the divalent and trivalent Fe cations residing in the 
R and I sites of the w\"ustite lattice are distinguishable by methods 
sensitive to changes in the electronic structure of valence shell 
(e.g. the M\"ossbauer spectroscopy), which are a consequence of changes 
in charge/spin states, electron localization/delocalization, and the defect structure. 
Indeed, the results of our calculations given in Table~\ref{Tab2} indicate stronger 
screening of $s$ electrons by 3$d$ electrons for the high-spin Fe$^{2+}$ ions compared 
to the high-spin Fe$^{3+}$ ions, which results in respectively higher and lower values
of the corresponding isomer shifts. Moreover, the isomer shift of the intersitial Fe$^{3+}$ 
is lower (0.384~mm/s) than the isomer shift of Fe$^{3+}$ occupying the R sites (0.487~mm/s). 
Similar behavior is encountered for Fe$^{2+}$ cations that exhibit slightly lower isomer 
shifts when located interstitially (1.039~mm/s) than octahedrally (1.108~mm/s). 
There is, however, no systematic dependence of $\delta$(Fe$^{2+}$) on the vacancy 
concentration $x$ contrary to $\delta$(Fe$^{3+}$) which slightly increases with increased $x$.
 
\begin{table}[b!]
\caption{Calculated ($x=0,3,6,9$\%) and experimental (bold numbers) 
\cite{McC85} isomer shifts of the Fe$^{2+}$ and Fe$^{3+}$ cations 
occupying the R and I sites of w\"ustite. 
Values of the isomer shifts for cations in 
the R positions are averaged over respective sites. Results are given 
for $U_{\mathrm{eff}}=5$~eV. According to the $\delta$ systematics 
\cite{Shenoy78}, compounds containing high-spin Fe$^{2+}$ show $\delta$ 
in the range of 1.0--1.5 mm/s, while for the compounds with high-spin 
Fe$^{3+}$ the $\delta$ ranges from 0.3 to 0.6 mm/s.}
\begin{ruledtabular}
\begin{tabular}{ccc}
 $x$ &$\delta$(Fe$^{2+}$) &$\delta$(Fe$^{3+}$)\\
 (\%) &(mm/s) &(mm/s) \\
\hline
0 &1.057 (R) &-- \\
\textbf{2.9} &\textbf{0.975--1.110} &\textbf{0.45} \\
3 &1.103 (R) &0.445 (R) \\
\textbf{4.8} &\textbf{0.933--1.005} &\textbf{0.60} \\
6 &1.097 (R) &0.453 (R) \\
9 &1.108 (R), 1.039 (I) &0.476 (R), 0.382 (I) \\ 
\end{tabular}
\end{ruledtabular}
\label{Tab2}
\end{table}

\begin{table}[t!]
\caption{The Bader charges and magnetic moments for different concentration of vacancies $x$.
The values of magnetic moments obtained from the FP-LAPW method are given in parentheses. 
The Bader charges and magnetic moments are averaged over respective sites.}
\begin{ruledtabular}
\begin{tabular}{cccccc}
    &   & \multicolumn{2}{c}{monovacancies} & \multicolumn{2}{c}{4:1 cluster} \\
 $x$ & $0$ & $3\%$ & $6\%$ & $5\%$ & $9\%$  \\
\hline 
\multicolumn{6}{c}{Bader charges ($e$)} \\
\hline
O$^{2-}$ & 7.32 & 7.28 & 7.25  & 7.28 & 7.26  \\
Fe$^{2+}$(R) &  6.68 & 6.70 & 6.71  & 6.69 & 6.69   \\
Fe$^{2+}$(I) &   -  &  - & - & 6.62 & 6.57 \\
Fe$^{3+}$(R) &   - & 6.40 & 6.38  & 6.33 & 6.33   \\
Fe$^{3+}$(I) &   -  &  - & - & 6.30 & 6.31 \\
\hline
\multicolumn{6}{c}{Magnetic moments ($\mu_B$)} \\
\hline 
Fe$^{2+}$(R) & 3.74 (3.68) & \multicolumn{2}{c}{3.67 (3.60)} & \multicolumn{2}{c}{3.69 (3.61)} \\
Fe$^{2+}$(I) &   -  & \multicolumn{2}{c}{-} & \multicolumn{2}{c}{3.72 (3.64)} \\
Fe$^{3+}$(R) &   - & \multicolumn{2}{c}{4.19 (4.17)} & \multicolumn{2}{c}{4.20 (4.18)} \\
Fe$^{3+}$(I) &   -  &  \multicolumn{2}{c}{-} & \multicolumn{2}{c}{4.15 (4.12)}
\end{tabular}
\end{ruledtabular}
\label{Tab3}
\end{table}

The calculated isomer shift of the interstitial Fe$^{3+}$ ion in Fe$_{1-x}$O is close 
to that measured for stoichiometric $\alpha$-$\mathrm{Fe_2O_3}$ (0.37-0.38~mm/s) \cite{Fe2O3} 
as well as that of the tetrahedral Fe$^{3+}$ ion in stoichiometric Fe$_3$O$_4$ 
at 4.2~K (0.39~mm/s) \cite{Fe3O4}. Furthermore, the isomer shift of Fe$^{3+}$ in 
the regular lattice sites of Fe$_{1-x}$O is comparable to the experimental value
of the octahedral Fe$^{3+}$ ion in Fe$_3$O$_4$ (0.51~mm/s). There is also a correspondence
between $\delta$(Fe$^{2+}$) in Fe$_{1-x}$O and $\delta$ of octahedral Fe$^{2+}$ in Fe$_3$O$_4$, 
since the M\"ossbauer resonance spectra of Fe$_3$O$_4$ obtained at 4.2~K reveal 
at least three components arising from the octahedral Fe$^{2+}$ ions, 
for which the isomer shifts range from 0.81 to 1.02~mm/s \cite{Fe3O4}. 
The sub-spectra due to the octahedral Fe$^{2+}$ ions results from the lowered symmetry 
of Fe$_3$O$_4$ below the Verwey transition temperature $T_V\approx120$~K. 
We mention that the room temperature M\"ossbauer spectra of Fe$_3$O$_4$ show 
the isomer shift for the octahedral cations of 0.66~mm/s resulting from the electron
hopping between the octahedral Fe$^{2+}$ and Fe$^{3+}$ ions. It is advantageous to compare 
the calculated and measured isomer shifts of Fe cations in Fe-doped Co$_{1-x}$O which is also 
a compound from the family of the simple 3$d$ transition-metal monoxides, albeit showing much 
less degree of non-stoichiometry (1--3\%) than w\"ustite. The high-spin Fe$^{2+}$ and Fe$^{3+}$ 
ions in Co$_{1-x}$O give rise to the M\"ossbauer resonance lines having respectively the isomer 
shifts of 1.02 and 0.37~mm/s \cite{Wdowik11}. These values remain in close relationship 
with those measured by the emission M\"ossbauer spectroscopy (1.12--1.14~mm/s for high-spin Fe$^{2+}$
and 0.34--0.37~mm/s for high-spin Fe$^{3+}$) \cite{CoO}. 

The M\"ossbauer spectra of realistic Fe$_{1-x}$O samples \cite{McC85,McC94} 
show much complexity as they are composed of several components 
reflecting quite a large number of different local environments 
accessible for Fe cations in the w\"ustite lattice. Although many 
attempts were undertaken \cite{MoessExp}, none of them described 
the local defect structure of w\"ustite in detail as the hyperfine 
parameters of the Fe cations deduced from the empirical fits could 
only provide a rough information on the statistically averaged 
environment of Fe$^{2+}$ and Fe$^{3+}$ in Fe$_{1-x}$O. 
Thus, the M\"ossbauer spectra of samples with the same stoichiometry 
could be successfully fitted using quite distinct phenomenological 
models. A difficulty in the interpretation of the experimental spectra 
presumably originates from a strong dependence of the w\"ustite defect 
structure on stoichiometry and the population of certain defect arrangements
in the Fe$_{1-x}$O lattice may vary with composition in a complicated manner. 
Although, the present theoretical research considers a limited number 
of possible defect configurations in w\"ustite, the results collected 
in Table~\ref{Tab2} can be useful while interpreting the M\"ossbauer 
spectra of Fe$_{1-x}$O because they provide direct relationship between 
the local electronic structure of Fe cations (reflected by the 
respective isomer shifts), their location inside the w\"ustite lattice, 
and their immediate surrounding. 

\begin{figure}[t!]
\includegraphics[width=\columnwidth]{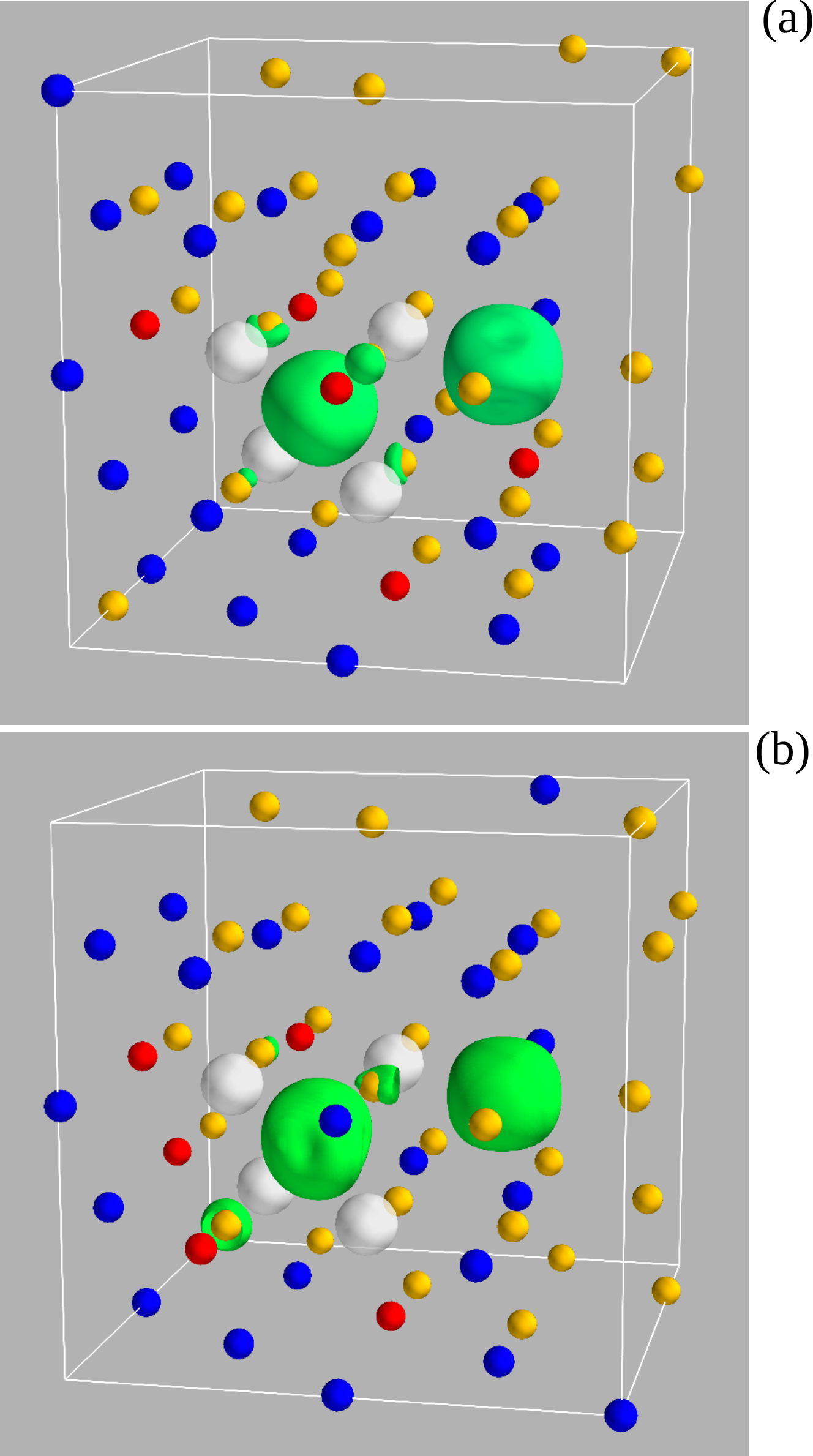} 
\caption{(Color online) Magnetization isosurfaces of Fe$^{2+}$ and 
Fe$^{2+}$ cations in Fe$_{1-x}$O containing 4:1-type vacancy clusters 
($x=9$\%) calculated at $U_{\mathrm{eff}}=5$~eV. (a) Configuration with 
the Fe$^{2+}$ ion at the R site and the Fe$^{3+}$ ion at the I site, 
(b) configuration with the Fe$^{2+}$ ion at the I site and the Fe$^{3+}$ ion 
at the R site. The cation vacancies and the R sites of Fe$^{3+}$ ions 
are denoted respectively by white and red spheres, whereas magnetization 
isosurface is marked in green. For clarity the isosurfaces are drawn only for 
the I site and one selected R site. Note relatively random distribution 
of Fe$^{3+}$ ions around the 4:1 defect.}
\label{Magnetization}
\end{figure}

We have calculated the charges of ions using the Bader analysis 
(see Table~\ref{Tab3}). Due to the hybridization effects, the obtained charges 
differ from the ionic valences and the difference between the Fe$^{2+}$
and Fe$^{3+}$ state is approximately 0.3 $e$. 
The trivalent cations exhibit enhanced spin magnetic moments comparing
to those of divalent ones. 
We note that the magnetic moments remain dependent on $U_{\mathrm{eff}}$ 
but weakly depends on $x$. Within $U_{\mathrm{eff}}$ ranging from 5 to 3 eV, 
they decrease linearly with decreasing $U_{\mathrm{eff}}$ 
by 0.085 and 0.048~$\mu_B$/eV for Fe$^{3+}$ and Fe$^{2+}$ cations, 
respectively. 
It is interesting to note that the magnetization isosurfaces around the 
Fe$^{2+}$ and Fe$^{3+}$ ions distinctly differ in their shapes, which 
can be seen in Fig.~\ref{Magnetization}. The isosurface of Fe$^{3+}$ 
is almost spherically symmetric, while that of Fe$^{2+}$ contains 
cavities. The symmetric distribution of magnetization reflects nearly 
equal occupation of five $3d$ orbitals in the Fe$^{3+}$ state, whereas 
one additional electron in the $t_{2g}$ orbital of Fe$^{2+}$ ion 
destroys such symmetric population and contributes to the visible 
asymmetry in the magnetization isosurface.  

\begin{figure}[t!]
\includegraphics[width=\columnwidth]{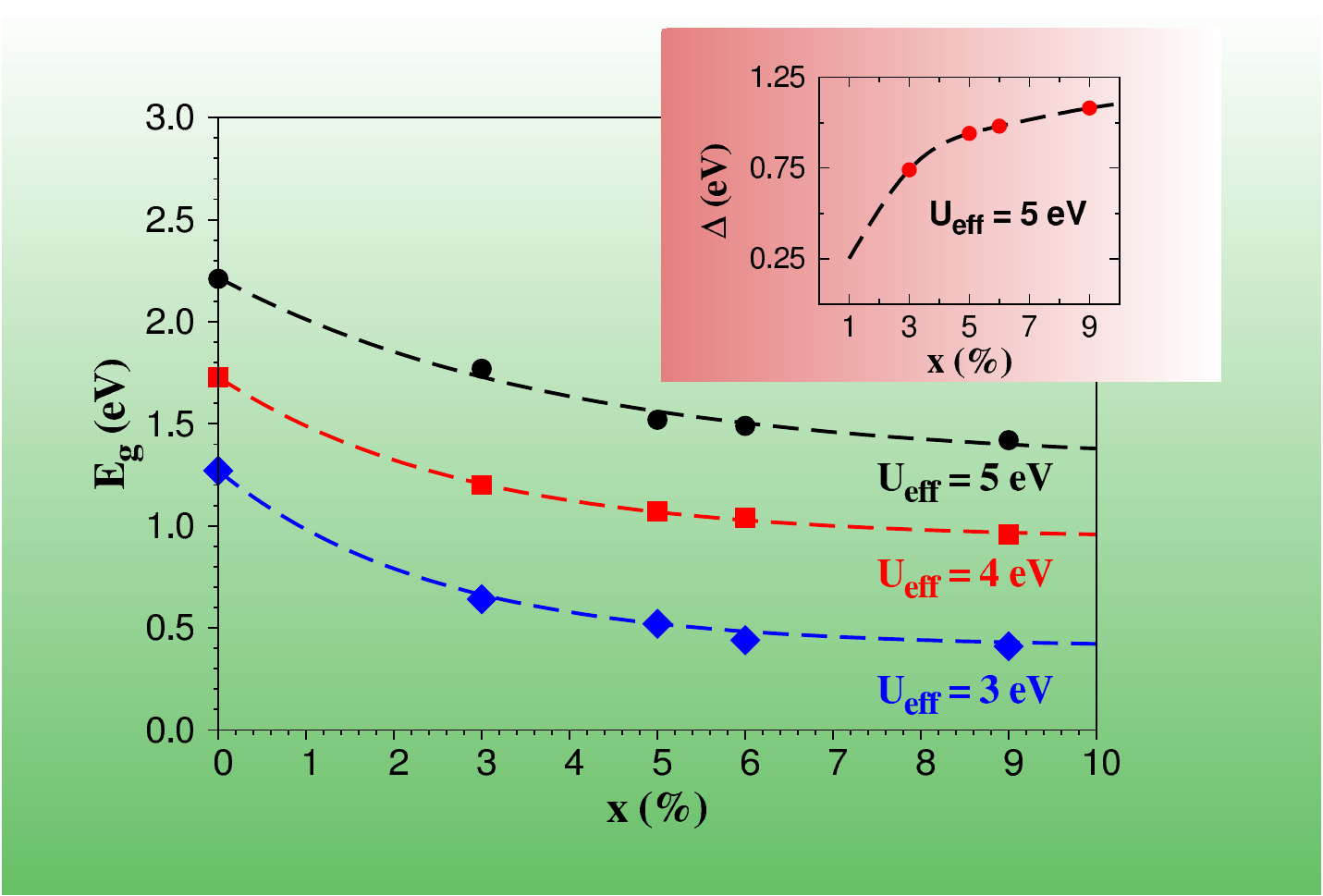}
\caption{(Color online) 
Influence of local interaction $U_{\mathrm{eff}}$ and vacancy 
concentration $x$ on the energy gap $E_g$ in Fe$_{1-x}$O. Inset shows 
the bandwidth ($\Delta$) of the empty states of Fe$^{3+}$ cations as a 
function of increasing $x$ for $U_{\mathrm{eff}}=5$~eV. 
Dashed lines are guides for an eye.}
\label{Gap}
\end{figure}

It is worth to analyze variation of the electronic properties of 
w\"{u}stite as a function of the Coulomb interaction parameter $U_{\mathrm{eff}}$. 
Although the exact value of $U_{\mathrm{eff}}$ remains unknown, 
it cannot be excluded that $U_{\mathrm{eff}}$ may depend on the defect 
concentration due to the screening processes induced by delocalized 
charge carriers. The effect of $U_{\mathrm{eff}}$ as well as the vacancy 
concentration on the band gap in Fe$_{1-x}$O are displayed in Fig.~\ref{Gap}. 
We observe that the energy gap of each Fe$_{1-x}$O 
composition decreases linearly upon decreasing $U_{\mathrm{eff}}$. 
Particular declines in the band gap energies are characterized by 
different slopes, i.e., the lower the $x$ the higher the slope. 
Obviously, the structures with higher vacancy concentrations exhibit 
more reduced energy gaps comparing to that of defect-free FeO, 
mainly due to the increased number of empty Fe$^{3+}$ states induced 
in the gap. These states are located just below the conduction band 
dominated by the unoccupied electronic states arising from the 
Fe$^{2+}$ ions that hybridize with the unoccupied states of oxygen 
anions (see Fig.~\ref{eDOS}). 
The best agreement with the experimental energy gap of 1.15~eV, measured 
for Fe$_{1-x}$O with $x=7$\% at 4.2~K \cite{Sch12}, is found 
for $U_{\mathrm{eff}}=4$~eV. The unoccupied trivalent states constitute 
a band having the width increasing nonlinearly with increasing vacancy 
concentration, as shown for typical $U_{\mathrm{eff}}=5$~eV in the inset 
of Fig.~\ref{Gap}. Similar behavior is also observed for the remaining values 
of $U_{\mathrm{eff}}$, although the lower values of the Hubbard interaction 
result in more pronounced broadening of that band.
Application of $U_{\mathrm{eff}}=2$~eV leads each Fe$_{1-x}$O 
composition to a metallic state and to disappearance of trivalent states in 
the w\"ustite lattice. In this case, all Fe cations become divalent 
with the average value of the spin magnetic moment of 3.57~$\mu_B$.

\section{Optical properties}
\label{sec:optic}

\subsection{Dielectric functions}
\label{sec:results}

In general, the frequency-dependent dielectric function of a solid is
a complex tensor, 
\begin{equation}
\varepsilon(\omega)=\varepsilon_1(\omega)+i\:\varepsilon_2(\omega). 
\label{eps}
\end{equation}
It can be used to characterize the linear response of a system to 
an electromagnetic radiation. The real $\varepsilon_1(\omega)$ and 
imaginary $\varepsilon_2(\omega)$ parts of $\varepsilon(\omega)$ 
describe dispersion and absorption of the radiation in a given material, 
respectively. 
In addition, the zero-frequency limit of $\varepsilon_1(\omega)$ 
corresponds to the electronic part of the static dielectric constant 
of a material $\epsilon_{\infty}$, a parameter being of fundamental 
importance in many aspects of material properties, whereas 
$\varepsilon_2(\omega)$ is more specific and remains closely related 
to the band structure of the system. 

The calculated dielectric function Eq. (\ref{eps}) of w\"ustite may 
exhibit anisotropic behavior due to the rhombohedral distortion induced 
by the AFII ordering and the removal of symmetry constraints arising 
from the presence of defects. Indeed, results of our calculations 
indicate subtle differences among the components of 
$\varepsilon(\omega)$ along the main crystallographic directions in the 
considered vacancy-defected systems of FeO. Nevertheless, one can 
neglect this effect as too tiny to be observed experimentally and 
consider the averaged values of $\varepsilon_1(\omega)$ and 
$\varepsilon_2(\omega)$. 

\begin{figure}[t!]
\includegraphics[width=\columnwidth]{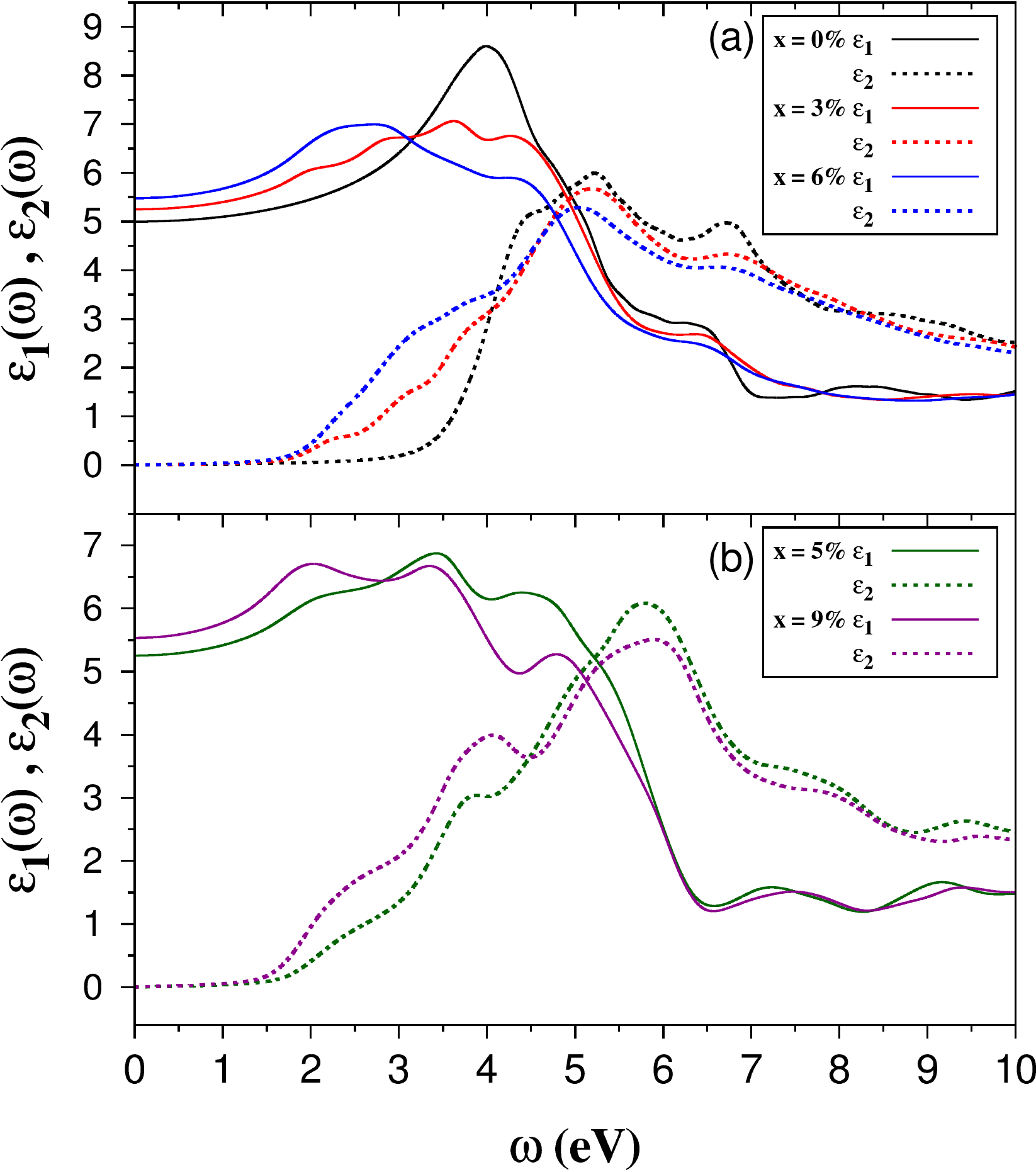}
\caption{(Color online) Real and imaginary parts of dielectric function 
(\ref{eps}) versus photon energy for (a) FeO with isolated vacancies, 
and (b) FeO with vacancy clusters. Calculations are performed with 
$U_{\mathrm{eff}}=5$~eV.}
\label{DielFun}
\end{figure}

Figure \ref{DielFun} shows that both components of $\varepsilon(\omega)$ 
remain sensitive to the vacancy concentration in the regime of low photon 
energies ($<8$~eV), while at higher energies the differences between 
dielectric functions in the systems with different $x$ are insignificant. 
Although $\epsilon_{\infty}$ increases with increased $x$, the shift 
upward does not exceed 10\% within the studied range of vacancy 
concentrations. The calculated values of $\epsilon_{\infty}$ at 
$U_{\mathrm{eff}}=5$~eV vary from 5.24 to 5.53 --- they correspond 
closely to the experimental $\epsilon_{\infty}=5.38$ given by Hofmeister 
\textit{et al.} \cite{Hof03} and theoretical value determined for 
perfectly stoichiometric FeO. On the other hand, one notes a broad
range of $\epsilon_{\infty}$ values deduced from a variety of 
experiments \cite{Sch12,Kant12,Seagle09,Prevot77,Kugel77}, which extends 
from 9.6 to 13. A considerably larger value of $\epsilon_{\infty}$ in 
Fe$_{1-x}$O in comparison to typical $\epsilon_{\infty}\approx 5$ for the 
remaining simple transition metal oxides (MnO, CoO, NiO) used to be 
ascribed to a large Fe deficiency reaching up to several percent 
\cite{Prevot77}. Much higher values of $\epsilon_{\infty}$ derived from 
experimental spectra may indicate an existence of large amount of 
delocalized charge carries in the highly nonstoichiometric samples that 
can be easily polarized enhancing the value of $\epsilon_{\infty}$. 
One should also be aware that the static dielectric constant is measured 
at finite temperature and involves a contribution arising from phonons, 
whereas the majority of the calculations are actually performed for a 
static crystal (zero temperature and neglecting zero-point vibrations) 
and hence they report values due to the purely electronic screening. 
This can be an additional source of errors, unless the phonon 
contribution is carefully extracted from the spectra.

The $\varepsilon_2(\omega)$ spectra clearly show the red shift of 
the absorption edge and reduction of the optical gap with increasing
vacancy concentration. The decrease of the optical gap from $\sim2.2$ 
to $\sim1.5$~eV while going from defect-free FeO to Fe$_{1-x}$O results 
from a presence of the Fe$^{3+}$ states in the band gap. A substantial 
reduction of the band gap in Fe$_{1-x}$O with vacancy content ranging 
from 7 to 8\% has been revealed by the recent infrared reflectivity and 
ellipsometry experiments \cite{Sch12,Park13}. These studies 
provide value of the fundamental absorption edge in w\"ustite of 1.15~eV 
at 5~K and 1.0--1.3~eV at room temperature. 

The absorption spectra of Fe$_{1-x}$O with different concentrations of 
vacancies show many common features. First of all, the spectra are 
dominated by a wide and intense peak at 5.2~eV which shifts slightly to lower 
(higher) energies for Fe$_{1-x}$O systems containing isolated vacancies 
(vacancy clusters). Another similarity is connected with the broad and less 
intense peak at 6.7~eV moving upward by about 1~eV but only in w\"ustite 
containing vacancy clusters. Generally, both peaks diminish 
their intensities with the increased defect content. The absorption spectra 
exhibit broad distribution in the energy range of 10--25~eV and 
structureless plateau declining progressively at still higher photon 
energies. 

There are, however, prominent differences in $\varepsilon_2(\omega)$ 
spectra between systems with isolated vacancies and 4:1-type clusters. 
They are encountered mainly below 4.5~eV and reflect differences 
in the electronic structures between those systems 
(cf. Fig.~\ref{eDOS} and Fig.~1 in Ref.~\cite{Wdowik13}). In the 
absorption spectra of w\"ustite containing isolated vacancies the 
shoulder extending from about 1.5 to 4~eV originates from interband 
transitions that involve states of Fe$^{3+}$ and Fe$^{2+}$ ions residing 
in R lattice positions as well as the oxygen states that are always 
hybridized with cation states. The respective feature in w\"ustite with 
vacancy clusters gets better resolved structure with a peak 
emerging at 4~eV and a tail with small swelling centered just above 2~eV. 
The 4~eV peak originates solely from the transitions due to the states 
of Fe$^{2+}$ ions at the R or I sites, provided the latter are present 
in the lattice [cf. densities of electronic states in Figs.~\ref{eDOS}(a) 
and \ref{eDOS}(b)]. 

On the other hand, the Fe$^{3+}$ states, either from R or I sites, that 
have the most considerable contribution inside the band gap, give rise 
to transitions observed at energy in the vicinity of 2~eV. It is worth 
to note that the swelling at 2~eV immerses into the broad distribution 
when Fe$^{3+}$ ions occupy only the R sites. In such a case the 
low-energy $\varepsilon_2(\omega)$ spectrum of w\"ustite with 4:1-type 
clusters resembles the $\varepsilon_2(\omega)$ spectrum of w\"ustite 
with isolated vacancies. Therefore, the swelling can be assigned to 
transitions involving trivalent Fe cations at interstitial lattice 
positions. Moreover, an exact identification of the transitions that 
are responsible for the peaks in the $\varepsilon_2(\omega)$ spectra 
of Fe$_{1-x}$O is hindered by a strong overlap and hybridization of 
the $s$, $p$, and $d$ states in both the valence and conduction bands. 
This effect has already been pointed out for perfectly stoichiometric 
FeO by R\"odl \textit{et al.} \cite{Rodl12}.

\subsection{Comparison with experiments}
\label{sec:exp}
 
\begin{figure}[t!]
\includegraphics[width=\columnwidth]{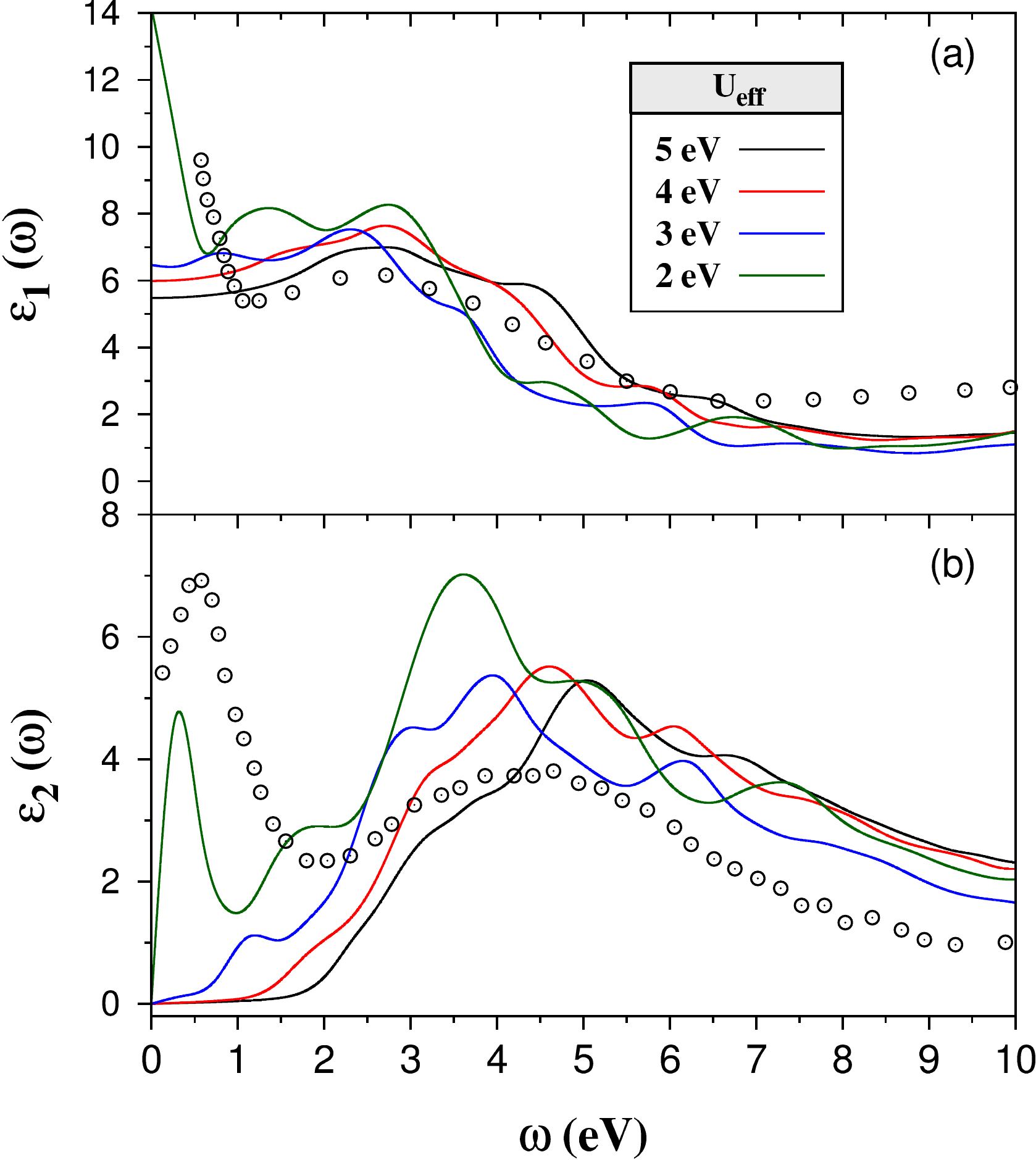}
\caption{(Color online) Dependence of 
(a) real and 
(b) imaginary parts of the dielectric function on 
$U_{\mathrm{eff}}$ for Fe$_{1-x}$O with $x=6$\%. 
The values derived from the experiment (open symbols) are 
adopted from Ref.~\cite{Hir09}.}
\label{DielFunU}
\end{figure}

Optical absorption spectra extracted from the reflectivity measurements 
\cite{Hir09} on w\"ustite with nonstoichiometry of the order of 
several percent show a very broad peak at about 4.5~eV and strongly 
enhanced absorption at low photon energies (see Fig.~\ref{DielFunU}). 
The latter is absent in our calculated $\varepsilon_2(\omega)$ spectra
obtained for $U_{\mathrm{eff}}=3-5$~eV. Some discrepancy between our 
theoretical results for $\varepsilon_2(\omega)$ and those derived from 
the experiment \cite{Hir09} remains unclear especially that both 
$\varepsilon_1(\omega)$ functions agree reasonably well in the line 
shape and magnitude within the energy range of 1--5~eV. We note, 
however, that the theoretical and experimental results are inconsistent 
below $\sim1$~eV due to abrupt growth of the experimental spectral 
amplitude of $\varepsilon(\omega)$. Such peculiar behavior of both 
$\varepsilon_1(\omega)$ and $\varepsilon_2(\omega)$ (not encountered in 
MnO, CoO, and NiO) was also observed in the data obtained from the 
ellipsometry spectroscopy \cite{Park13} and has been attributed to 
defect absorption at Fe vacancies. Indeed, the experimental 
$\varepsilon_1(\omega)$ and $\varepsilon_2(\omega)$ spectra exhibiting 
growth with $\omega \rightarrow 0$ seem to reflect the narrow-gap nature 
of Fe$_{1-x}$O at vacancy concentrations of the order of several percent. 

The increase in absorption spactra at low energies, typical for
metals (Drude peak), indicates the existence of delocalized
charge carriers despite of the Mott insulator state of w\"{u}stite.
These itinerant carriers (electrons or holes) exist due to the defected 
structure of Fe$_{1-x}$O. Such a complicated electronic structure is 
difficult to model within the DFT approach. However, the effect of 
w\"ustite \textit{metallization} can be to some extent simulated by  
lowering the Hubbard parameter $U$ on the Fe cations. 

Results of such investigations are depicted in Fig.~\ref{DielFunU}. 
One observes shifting of the peak's positions to lower energies with decreasing 
$U_{\mathrm{eff}}$ from 5 to 4~eV. At still lower $U_{\mathrm{eff}}$ 
the spectra acquire more complex pattern composed of additional peaks 
appearing at low $\omega$, and finally the Fe$_{1-x}$O system transforms 
to a metallic state at $U_{\mathrm{eff}}=2$~eV .
The $\varepsilon_2(\omega)$ spectrum shows Drude-like peak at low energies
in good agreement with the data derived from the experiment. 
The \textit{metallization} arising from its large off-stoichiometry also 
affects the real part of the dielectric function which displays 
pronounced growth at low $\omega$, which in turn yields substantial 
increase of the electronic part of static dielectric constant 
($\epsilon_{\infty}$=14.2). This phenomenon is likely to be responsible 
for the characteristic features of w\"ustite dielectric functions 
detected in a variety of experiments and commonly interpreted as 
defect-related effects.            

\begin{figure}[t!]
\includegraphics[width=\columnwidth]{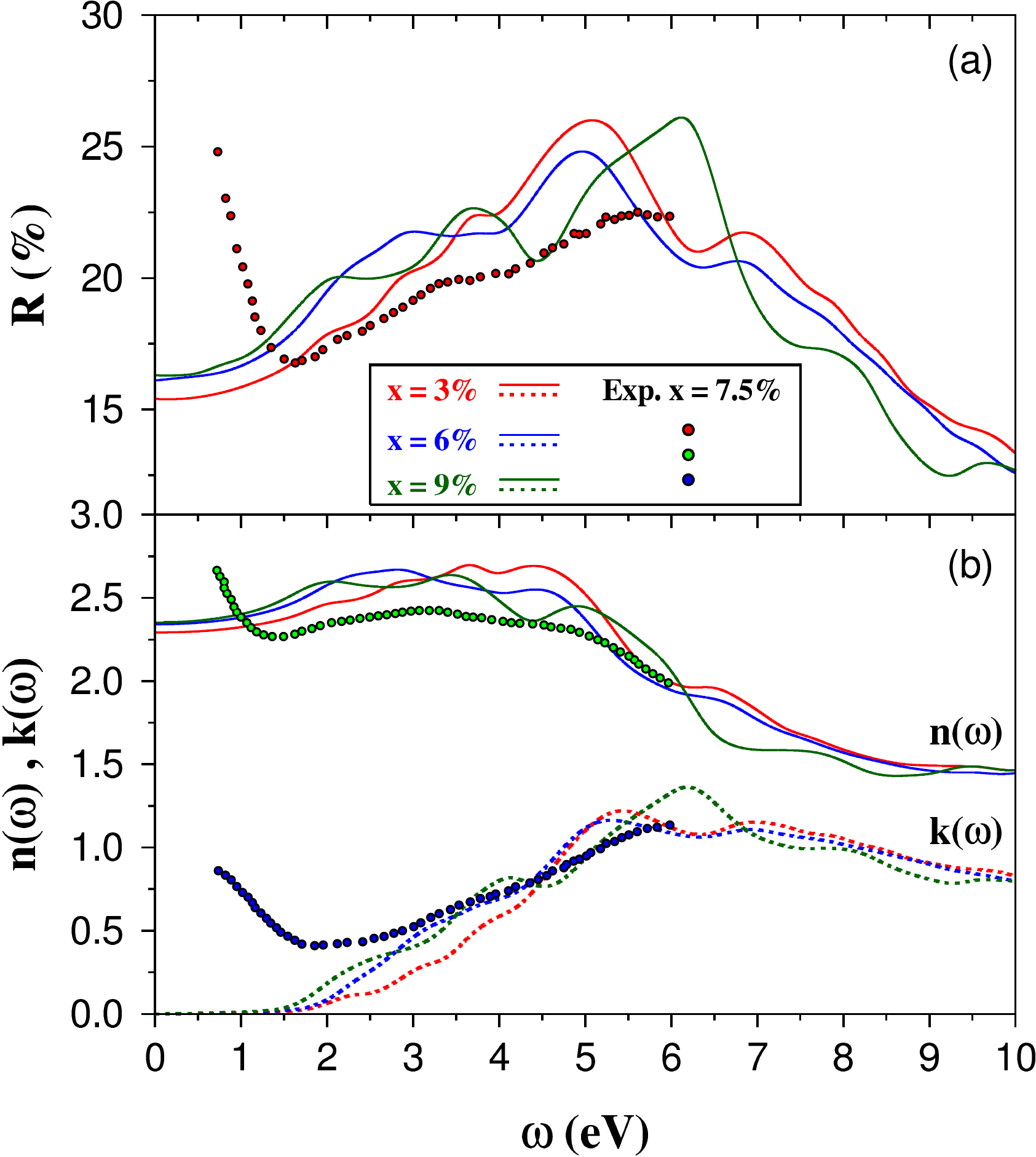}
\caption{(Color online)
(a) Reflectivity $R(\omega)$, (b) refractive index $n(\omega)$  
and extinction coefficient $k(\omega)$ of Fe$_{1-x}$O with different 
nonstoichiometry. Theoretical and experimental data \cite{Park13} 
are denoted by solid/dashed lines and symbols, respectively. Theoretical
results are obtained for $U_{\mathrm{eff}}=5$~eV.}
\label{Rnk}
\end{figure}

The real and imaginary components of $\varepsilon(\omega)$ are usually 
determined from ellipsometry experiments or derived from absorption, 
reflection, or transmission measurements after the Kramers--Kronig 
analysis. The experimental data are frequently presented in the form of 
the reflectivity spectrum \cite{Fox01}, 
\begin{eqnarray}
R(\omega) &=& \left| 
\frac{\sqrt{\varepsilon(\omega)}-1)}{\sqrt{\varepsilon(\omega)}+1)} 
\right|^2 =\frac{\left[n(\omega)-1\right]^2+k^2(\omega)}{\left[n(\omega)
+1\right]^2+k^2(\omega)}, 
\end{eqnarray} 
where 
\begin{eqnarray}
n(\omega)&=&\sqrt{\frac{|\varepsilon(\omega)|+\varepsilon_1(\omega)}{2}},\\
k(\omega)&=&\sqrt{\frac{|\varepsilon(\omega)|-\varepsilon_1(\omega)}{2}}.
\end{eqnarray} 
The functions $n(\omega)$ and $k(\omega)$ denote respectively refractive index 
and extinction coefficients that constitute the complex refractive index, 
\begin{equation} 
N(\omega) = \sqrt{\varepsilon(\omega)} = n(\omega) + i k(\omega). 
\end{equation} 

A comparison between results of calculations performed as a function of 
$x$ and the recent ellipsometry experiments carried out on samples 
with $x=7.5$\% \cite{Park13} is given in Fig.~\ref{Rnk}.
Similarly to the dielectric functions,  
the calculated spectra shift to lower energies for larger concentration 
of defects. Thus, the zero energy limit of the reflectivity and 
the refractive index increases and the absorption edge visible in 
the extinction coefficient diminishes with the increasing $x$.
Despite small differences, one  finds that the calculated 
$R(\omega)$, $n(\omega)$, and $k(\omega)$ spectra
reasonably well reflect the behavior of the respective experimental 
quantities above $\sim 1.5$~eV. Both experimental and calculated data 
indicate an increase in the reflectivity of Fe$_{1-x}$O by almost 10\% 
between 1.5--5.0~eV and the reflectivity decrease above this energy 
range. A visible mismatch between theoretical and experimental results 
is observed at low photon energies, where the measured quantities show 
pronounced increase, probably due to higher amount of free charge 
carriers embedded in the measured samples as well as much stronger 
absorption by defects in comparison with the simulated Fe$_{1-x}$O 
systems.

\begin{figure}[t!]
\includegraphics[width=\columnwidth]{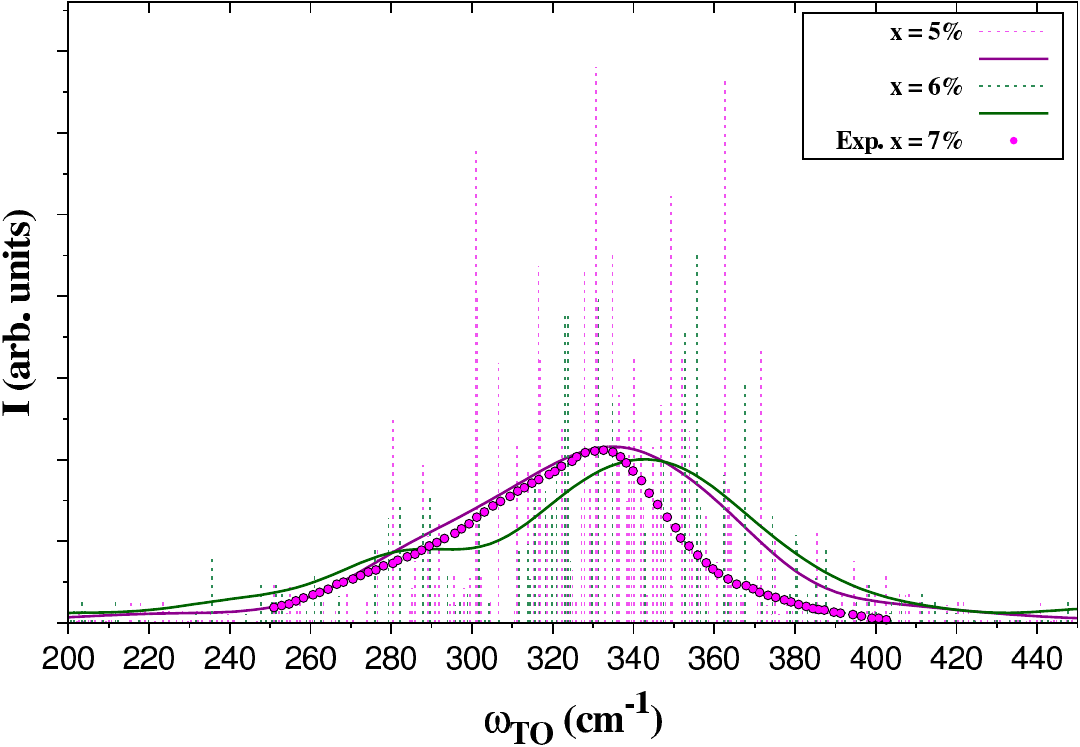}
\caption{(Color online) Theoretical infrared absorption intensities 
(dotted vertical lines) of Fe$_{1-x}$O with the Fe-vacancy content of 
5\% (vacancy clusters) and 6\% (isolated vacancies) and the Gaussian 
convolutions of theoretical spectra with FWHM of 33 $\mathrm{cm}^{-1}$ 
(solid lines). Experimental dielectric loss function \cite{Sch12} 
measured at 5~K for Fe$_{1-x}$O with $x=7-8$\% is represented by solid 
symbols. The frequencies $\omega_{\mathrm{TO}}$ correspond to the 
transverse optic phonons. Theoretical results are obtained for 
$U_{\mathrm{eff}}=5$~eV.}
\label{infas}
\end{figure}

A deficiency in the w\"ustite cation sublattice was also suggested 
to be responsible for large spectral linewidth of the transverse 
optic (TO) phonon obtained from the infrared reflectivity measurements 
\cite{Sch12,Kant12}. 
One finds that the dielectric loss spectrum of the 
paramagnetic Fe$_{1-x}$O contains broad and symmetric single line 
peaked at about 325~$\mathrm{cm}^{-1}$, whereas the antiferromagnetic  
Fe$_{1-x}$O reveals weakly resolved two-peak structure making the 
low-temperature experimental spectrum asymmetric. This feature is 
also observed in our Gaussian convoluted infrared absorption spectrum, 
which is directly related to dielectric loss function. Indeed, as 
depicted in Fig.~\ref{infas}, both theoretical and experimental 
infrared absorption spectra show much broader distributions of the TO 
phonon frequencies than those in more stoichiometric simple $3d$ 
transition metal oxides (MnO, CoO, and NiO)~\cite{Kant12}.
 
The broad theoretical spectrum arises from the splitting of phonon 
frequencies due to the crystal symmetry being lowered by the presence 
of defects. The $\delta$-functions, shown in Fig.~\ref{infas}, 
represent the infrared intensity $\delta$-peaks associated with the 
splitted frequencies, subsequently convoluted with the experimental 
Gaussian width. 
It appears that the maximum of the convoluted spectral peak for the 
simulated compositions of $x=5$ and 6\% shifts to higher frequencies 
(330 and 342~$\mathrm{cm}^{-1}$) in comparison with the position of 
the respective peak in the defect-free FeO (324~$\mathrm{cm}^{-1}$). 
Also, the main peak in the Gaussian convoluted infrared absorption 
spectra changes its position from 338~$\mathrm{cm}^{-1}$ to 
319~$\mathrm{cm}^{-1}$, while increasing $x$ from 3 to 9\% (not shown).

A similar trend to the one observed here 
has already been encountered in the early infrared 
reflectivity experiments \cite{Bowen75} performed on w\"ustite samples 
with different concentrations of cation vacancies. 
Such a strong sensitivity of the infrared spectra to the concentration 
of defects can account for a large scattering seen in the experimentally 
determined TO phonon frequencies in w\"ustite which cover the range 
extending from 320 to 410~$\mathrm{cm}^{-1}$  
\cite{Sch12,Kant12,Seagle09,Hof03,Kugel77}. The broadening and shifting 
of the infrared spectral peaks seem to have the same origin as very similar 
changes observed in the phonon density of states \cite{Wdowik13} and  
they reflect pronounced modifications of the electronic structure and 
force constants due to cation vacancies in the Fe-sublattice of w\"ustite.    

\section{Summary and conclusions}
\label{sec:sum}

This work investigates the role of strong electron correlations on Fe 
atoms and the high concentration of cation vacancies in modifying 
the electronic and dielectric properties of w\"{u}stite 
(Fe$_{1-x}$O). As we have shown, both of them influence 
substantially the electronic properties of this compound and have 
opposite effect on its band structure. While the local electron 
interactions in the Fe $3d$ states are responsible for the opening of 
the insulating gap, the Fe vacancies reduce its magnitude. The 
mechanism of gap reduction remains the same, irrespectively of the 
type of incorporated defects. Either monovacancies or vacancy 
clusters induce the band of empty Fe$^{3+}$ states inside the gap of 
the perfect FeO crystal. The width of this band increases monotonically 
with the increased concentration of cation vacancies reducing the distance 
from the top of valence states, so effectively diminishing the band gap.

The strength of the effective local interaction $U_{\textrm{eff}}$ 
depends presumably on the vacancy content, i.e., the band gap reduction 
enhances the screening of Coulomb interactions, which may lead to even 
larger electron mobility and {\it metallization} of iron oxide. 
The presence of free charge carriers, modeled here by small values 
of $U_{\mathrm{eff}}$, explain the increase of dielectric function and 
optical reflectivity observed experimentally at low energies. 

Summarizing, we presented an explanation of several effects observed in 
the optical properties of w\"{u}stite and explained them by the 
presence of defects.
We emphasize that our studies also explain the anomalous 
broadening of infrared spectra, demonstrating a strong effect of cation 
vacancies on phonons in w\"{u}stite. Similar correlations between  
charge distribution and lattice dynamics induced by the electron-phonon 
coupling play an important role in the Verwey transition in magnetite 
(Fe$_3$O$_4$) \cite{Verwey}. 
As noticed recently \cite{Ber14}, also the polaronic 
distribution found in defected FeO resembles the charge-orbital order 
in magnetite and may share a common origin with the short-range
order observed above the Verwey temperature \cite{Bosak}.
Finally, we point out that the effects discussed here can be relevant 
not only for iron oxides but are expected to determine the electronic, 
dielectric, and dynamical properties of other strongly correlated 
systems with high concentrations of intrinsic defects as well.

\acknowledgments
The authors acknowledge valuable discussions with Ma\l{}gorzata Sternik 
and Jan \L{}a\.{z}ewski, Institute of Nuclear Physics, 
Polish Academy of Sciences, Cracow, Poland.  
Interdisciplinary Center for Mathematical and Computational Modeling 
(ICM), Warsaw University, Poland and the IT4Innovations National 
Supercomputing Center, VSB-Technical University, Ostrava, Czech 
Republic are acknowledged for providing the computer facilities 
under Grants No. G28-12 and Reg. No. CZ.1.05/1.1.00/02.0070.
The authors acknowledge support by the Polish National Science Center 
(NCN) under Projects No. 2011/01/M/ST3/00738 and No. 2012/04/A/ST3/00331.

\end{document}